\documentclass[a4paper]{PoS}

\title{Radiative charged-lepton masses with more than one Higgs doublet}

\ShortTitle{Radiative charged-lepton masses with more than one Higgs doublet}

\author{Filipe Joaquim$^a$ and \speaker{Jo\~ao Penedo}$^{ab}$\\
\llap{$ˆa$} CFTP, Departamento de F\'isica, Instituto Superior T\'ecnico,\\
Universidade de Lisboa, Avenida Rovisco Pais 1, 1049 Lisboa, Portugal\\
\llap{$ˆb$} SISSA, Via Bonomea 265, I-34136 Trieste, Italy\\
        E-mail: \email{filipe.joaquim@tecnico.ulisboa.pt}\\
        \email{joao.t.n.penedo@tecnico.ulisboa.pt}}

\abstract{%
We discuss how charged-lepton masses can be induced in multi-Higgs doublet models (NHDMs) through renormalization group running of Yukawa couplings. Examples of electron and muon mass generation are presented within scenarios with two- and three-Higgs-doublet models. We also show that quantum corrections to the Yukawa couplings can be naturally of the same order as the tree-level values. The impact of such corrections in NHDMs with right-handed neutrinos is briefly commented.\footnote{This contribution is based on the work of Ref.~\cite{Joaquim:2014gba}.}}

\FullConference{Proceedings of the Corfu Summer Institute 2014 \\

                 3-21 September 2014\\

                 Corfu, Greece}

\usepackage{cancel}
\usepackage{graphicx}
\usepackage{epsfig}
\usepackage{dsfont}
\usepackage{amsmath,amsfonts,amssymb}
\usepackage{t1enc}

\newcommand{\bY}{{\rm \bf Y}}
\newcommand{\bM}{{\rm \bf M}}
\newcommand{\bV}{{\rm \bf V}}
\newcommand{\bU}{{\rm \bf U}}
\newcommand{\phia}{\Phi_a}
\newcommand{\bA}{{\rm \bf A}}
\newcommand{\bB}{{\rm \bf B}}
\newcommand{\bX}{{\rm \bf X}}

\begin{document}

\section{Introduction}
The recent discovery of a Higgs-like boson at the Large Hadron Collider (LHC)~\cite{Aad:2012tfa} represents an important milestone on the understanding of mass generation and electroweak symmetry breaking (EWSB). Nevertheless, an explanation for the observed distribution of fermion masses is lacking. Hierarchies in this pattern suggest that while third-generation masses may arise at the classical level, quantum corrections are responsible for the origin of the remaining ones \cite{'tHooft:1971rn}. Realistic models of radiative fermion mass and mixing generation have been built both in the frameworks of supersymmetric theories~\cite{Nanopoulos:1982zm} and Grand Unified Theories (GUTs)~\cite{Barr:1979xt}.
Fermion mass hierarchies may alternatively be explained by the presence of symmetries acting on flavor space and constraining the structure of the Yukawa couplings. One can consider, for instance, the Froggatt-Nielsen scenario~\cite{Froggatt:1978nt} where U(1) shaping symmetries are broken when some scalar fields $S$ ({\em flavons}) acquire nonvanishing vacuum expectation values (VEVs), $\langle S \rangle$. The effective Yukawa structures thus generated are governed by powers of small parameters $\epsilon \sim \langle S \rangle / M$.

Concerning the explanation of the neutrino flavour pattern, where the mass hierarchy is mild and the mixing large, the recently taken route is to implement discrete symmetries~\cite{Altarelli:2010gt}. One possible model-building approach in this context is to extend the field content by adding several scalar doublets, whose couplings to fermions depend on the broken or unbroken symmetries, and may or not be generated at the renormalizable level. The VEV and Yukawa coupling configurations will then determine the pattern of fermion masses. If a certain VEV configuration cannot accommodate the observed values of fermion masses at tree level, it is customary to dismiss it as not phenomenologically viable. This would occur if, for instance, a charged lepton $e_i$ is massless (or too light) at tree level in a multi-Higgs doublet model where only one of the doublets carries a VEV. In this case, a sizable coupling of these leptons with the nonzero-VEV Higgs can nevertheless be induced by quantum corrections, contributing to a mass $m_{e_i}$ after EWSB. In the present work, the possibility of radiatively generating charged-lepton masses due to such corrections is explored. Simple examples in two- and three-Higgs doublet models are considered, and the case of NHDMs extended with right-handed neutrinos is also briefly discussed.

\section{Radiative charged-lepton masses in NHDM\lowercase{s}}
Consider the extension of the Standard Model (SM) with $N$ Higgs doublets $\phia=(\phi_a^+,\phi_a^0)^T \sim (2,1/2)$. The Yukawa Lagrangian reads
\begin{equation}
\label{Lql}
-\mathcal{L}=(\bY_a^u)_{ij} \bar{q}_{Li} \tilde{\Phi}_a u_{Rj} +(\bY_a^d)_{ij} \bar{q}_{Li} \phia d_{Rj}  +(\bY_a^\ell)_{ij} \bar{\ell}_{Li} \phia e_{Rj} +{\rm H.c.}\,,
\end{equation}
where $q_{Li}$, $\ell_{Li}$ denote quark and lepton doublets, $u_{Ri}$, $d_{Ri}$ and $e_{Ri}$ correspond to the right-handed singlets, and $\tilde{\Phi}_a=i\sigma_2 \phia^\ast=(\phi_a^{0\ast},-\phi_a^{-})^T$. The Yukawa matrices $\bY_a^X$ are general $3\times 3$ complex matrices which can be diagonalized by the biunitary transformations
\begin{equation}
\label{Biunit}
\bV_a^{X\dag} \bY_a^X \bU_a^X={\rm diag}(y_{a1}^X,y_{a2}^X,y_{a3}^X)\,,
\end{equation}
with $y_{ai}^X$ are real and positive. After EWSB, mass matrices are generated at tree level as
\begin{equation}
\label{Ms}
\bM_{d,\ell}=\sum_{a=1}^N v_a \bY_a^{d,\ell}  \; , \;
\bM_{u}=\sum_{a=1}^N v_a^\ast \bY_a^u\,,
\end{equation}
where $v_a=\langle \phi_a^0 \rangle$ is the VEV of $\phi_a^0$. We restrict our attention to corrections to the charged-lepton Yukawa couplings, $\bY_a^\ell$. At the one-loop level, the renormalization group equations (RGEs) for these couplings read
\begin{equation}
\label{RGE}
16 \pi^2 \frac{d \bY_a^\ell}{dt}=\beta_{a}\;,\; t=\log\left(\frac{\mu}{\Lambda}\right)\,,
\end{equation}
with $\mu$ and $\Lambda$ the renormalization and reference energy scales, respectively. For the NHDM, the beta function at one loop is~\cite{Grimus:2004yh} 
\begin{equation}
\label{beta}
\beta_{a}^{(1)}=\alpha_g \bY_a^\ell+\alpha_Y^{ab} \bY_b^\ell+\bY_a^\ell \bY_b^{\ell\dag} \bY_b^\ell+\frac{1}{2} \bY_b^\ell \bY_b^{\ell\dag} \bY_a^\ell\,,
\end{equation}
with $\alpha_g=-9g^2/4-15g^{\prime 2}/4$, where $g$ and $g^\prime$ are the SU(2)$_W$ and U(1)$_Y$ gauge couplings and 
\begin{equation}
\label{alphas}
\alpha_Y^{ab}=3{\rm Tr}(\bY_a^{u\dag} \bY_b^u+\bY_a^d\bY_b^{d\dag} )+{\rm Tr}(\bY_a^\ell \bY_b^{\ell\dag} )\,.
\end{equation}
The contribution controlled by these coefficients stems from Higgs wave-function diagrams with ingoing $\Phi_a$, quarks and leptons in the loop, and outgoing $\Phi_b$.

In order to simplify our analysis, we henceforth consider the Higgs basis~\cite{Lavoura:1994fv} (although the analysis is valid for a general vacuum configuration) where all VEVs are zero except the one of some $\phi_a^0$ ($v_a=v=174$~GeV), and Yukawa couplings are rotated accordingly. The charged-lepton mass matrix is therefore
\begin{equation}
\label{Mell}
\bM_\ell=v \bY_a^\ell\;,\; \bV_a^{\ell\dag} \bY_a^\ell \bU_a^\ell={\rm diag}(y_e,y_\mu,y_\tau)
\end{equation}
at tree level. 
Notice that we are not interested here in possible flavor-changing neutral current constraints~\cite{Mahmoudi:2009zx}, since these can be avoided considering the decoupling limit of NHDMs~\cite{Haber:1989xc}. 

We will focus in the one-loop corrected couplings, denoted by $\bY_a^{\ell (1)}$. The beta function for $\bY_a^\ell$ given in Eq.~\eqref{beta} contains terms depending on $\bY_a^\ell$ itself, contributing to existing tree-level masses, and terms proportional to $\bY_b^\ell$, namely the aforementioned $\alpha_Y^{ab} \bY_b^\ell$ term. For $b\neq a$, such terms induce corrections to $\bY_a^\ell$ which are independent from its own structure. Under the reasonable assumption that the top quark couples with the same strength to all Higgs doublets, $y_t\simeq 1$, and considering, for simplicity, all matrices $\bY_b^\ell$ diagonal with elements given by $y_{bi}$, one can estimate the magnitude of corrections to $\bY_a^\ell$ coming from $\bY_b^\ell$ to be typically of the order of 
\begin{equation}
\label{deltay}
\delta y_{ai}\sim  \frac{3 y_{bi}}{16\pi^2}\log\left(\frac{\Lambda}{m_H}\right) \;,\; i=1,2,3\;,\; b\neq a\,,
\end{equation}
where $\Lambda$ is a high scale at which the Yukawa couplings are initially given, and $m_H$ is the typical scale of the extra scalars in the theory. This illustrates the fact that Yukawa couplings with the zero-VEV scalars, while not contributing to tree-level masses, can induce important corrections to $\bY_a^\ell$, depending on $\Lambda$, $m_H$ and on the size of the $\bY_b^\ell\,(b\neq a)$.
In order to end up with three massive charged leptons after taking quantum corrections into account, one requires $r(\bY_a^{\ell (1)})=3$, where $r$ denotes matrix rank. It can be seen that, in general, the rank of Yukawa matrices in a NHDM may change due to these corrections, and lepton masses which were absent at tree level are induced. One starts by noticing that the beta function for $\bY_a^\ell$, at any loop order $n$, can be cast in the form
\begin{equation}
\label{betaform}
\beta_a^{(n)} = \sum_{k=1}^N \, \bX_{a,k}^{(n)}\, \bY_k^\ell\, ,
\end{equation}
where $\bX_{a,k}^{(n)}$ are matrices in flavor space. For instance, from Eq.~\eqref{beta}, one has at one loop
\begin{equation}
\label{X1loop}
\bX_{a,k}^{(1)} = \big(\delta_{ak}\,\alpha_g+\alpha_Y^{ak}\big)\,\mathds{1} + \bY_a^\ell {\bY_k^\ell}^\dagger + \frac{1}{2}\, \delta_{ak}\, \bY_a^\ell {\bY_a^\ell}^\dagger\,.
\end{equation}
In the leading-log approximation, the structure of the couplings $\bY_a^{\ell (n)}$ (corrected up to order $n$) will be given in the form of Eq.~\eqref{betaform} with different $\bX$ matrices, $\bX_{a,k}^{\prime(n)}$. In the general case where at least one $\alpha_Y^{ab}$ ($b\neq a$) is nonzero, maximum rank for these $\bX^\prime$ matrices is expected already at one-loop level.
Since $r(\bA \bB)\leq \min\{r(\bA),r(\bB)\}$, it is clear that the rank of each term in the sum of Eq.~\eqref{betaform} is at most equal to $r(\bY_k^\ell)$, while since rank is subadditive, {\em i.e.} $r(\bA+\bB) \leq r(\bA)+ r(\bB)$. One thus has
\begin{equation}
\label{rankbeta}
r(\bY_a^{\ell (n)})=r\Bigg( \sum_{k=1}^N \, \bX_{a,k}^{\prime(n)}\, \bY_k^\ell\Bigg) \, \, \leq \, \, {\rm min}\left\{3,\sum_{k=1}^N \, r(\bY_k^\ell)\right\}\,.
\end{equation}
Barring tailored cancellations among Yukawa structures, equality generally holds in Eq.~\eqref{rankbeta}. This result implies that one can in principle generate charged-lepton masses from low-rank tree\discretionary{-}{-}{-}level Yukawa couplings. Hence, the number of massive charged leptons will depend on these ranks and on the number of existing Higgs doublets. For instance, to end up with $n_m$ massive charged leptons from rank-1 (tree-level) mass and Yukawa matrices one would need\footnote{This condition is necessary but not sufficient due to the aforementioned possibility of cancellations, a trivial such exception being $\bY_2^\ell \propto \bY_1^\ell$.} $N \geq n_m$. To make this point clear, we consider in what follows some simple and extreme examples where radiative mass generation is key. It is not our goal to frame these examples in specific models but to provide a proof of concept for these claims.

Considering first the 2HDM case~\cite{Branco:2011iw}, one can see that at most one charged-lepton mass can be radiatively generated from rank-1 Yukawa matrices in the Higgs basis. Instead, if $3-n$ charged leptons are massive at tree level, one needs $r(\bY_2^\ell) \geq n$ to end up with all charged leptons massive.

In the 3HDM, on the other hand, all three charged leptons could acquire a mass from rank-1 Yukawa matrices in the Higgs basis. To illustrate this, consider a 3HDM with a vacuum configuration of the type $(v_1,v_2,v_3)=(0,0,v)$ and with Yukawa couplings\footnote{Notice that in the following examples the zero entries in the Yukawa matrices should perhaps not be taken as strict zeros but interpreted as the limit of very suppressed entries. Our results are valid also for more complicated flavor structures which we do not consider here.} $\bY_1^\ell={\rm diag}(\epsilon_1,0,0)$, $\bY_2^\ell={\rm diag}(0,\epsilon_2,0)$, $\bY_3^\ell={\rm diag}(0,0,\epsilon_3)$, taking all parameters real thus neglecting possible CP-violating effects in this sector~\cite{Branco:2011zb}. Within this setup, only the tau is massive at tree level and only $\bY_3^\ell$ is relevant for charged-lepton masses. At one loop, in the leading-log approximation, the charged-lepton masses thus obtained read
\begin{equation}
\label{Mlex}
m_{e_{i}}^{(1)}\simeq \frac{v}{16 \pi^2}\alpha_Y^{3i} \epsilon_i \log\left( \frac{\Lambda}{m_H}\right)\;,\; i=1,2\;.
\end{equation}
Hereafter we consider $\Lambda = \Lambda_{\rm GUT} \sim 10^{16}$~GeV and $m_H \sim 1$~TeV. Keeping $\alpha_Y^{3i} \sim \mathcal{O}(1)$, the right masses are in this case obtained for $(\epsilon_1,\epsilon_2,\epsilon_3)\simeq(1.6\times 10^{-5},3.2\times 10^{-3},0.01)$, whose hierarchical structure could be explained by a Froggatt-Nielsen mechanism.

Even though so far only extreme examples of mass generation have been considered, less trivial Yukawa structures can also introduce corrections to lepton mixing. This effect can be illustrated by considering the same 3HDM as before with a nondiagonal choice of rank-1 $\bY_a^\ell$ matrices:
\begin{align}
\label{Ysexam}
&\bY_1^\ell=
\left(
\begin{array}{ccc}
\epsilon_1 & 0 &0 \\
- \epsilon_1 \epsilon  &0 &0 \\
- \epsilon_1\epsilon  &0 &0 \\
\end{array}
\right)\;,\;
\bY_2^\ell=\left(
\begin{array}{ccc}
0 &\epsilon_2\epsilon   & 0  \\
0 & \epsilon_2 &0 \\
0 &0 &0
\end{array}
\right)\;,\;
\bY_3^\ell=\left(
\begin{array}{ccc}
0 &0  & \epsilon_3 \epsilon   \\
0 & 0 & 0 \\
0 &0 &\epsilon_3
\end{array}
\right)\,.
\end{align}
As before, only one charged lepton is massive at tree level, while taking radiative effects into account corrects $\bY_3^\ell$ to $\bY_3^{\ell (1)}=\bY_3^{\ell}-\delta\bY_3^{\ell}$, with
\begin{align}
\label{Y3ex}
\delta\bY_3^{\ell}\simeq \frac{1}{16\pi^2}
\left(
\begin{array}{ccc}
\alpha_Y^{31}\epsilon_1 & {\alpha_Y^{32} \epsilon_2\epsilon}  &\dfrac{1}{2}{\epsilon_3 \epsilon(\epsilon_2^2 \epsilon^2+6\epsilon_3^2)} \medskip\\
- {\alpha_Y^{31}\epsilon_1 \epsilon}  &\alpha_Y^{32}\epsilon_2 &\dfrac{1}{2}\epsilon_2^2\epsilon_3\epsilon^2 \medskip\\
- {\alpha_Y^{31}\epsilon_1 \epsilon}   & 0 &\dfrac{3}{2}\epsilon_3^3\\
\end{array}\right)\log\left(\dfrac{\Lambda}{m_H}\right)\,,
\end{align}
and $\alpha_Y^{3i}$ given by Eq.~(\ref{alphas}). At this (one-loop) level, one obtains
\begin{align}
\label{msexa}
m_{e}^{(1)}\simeq \frac{\alpha_Y^{31}\epsilon_1}{16\pi^2} \sqrt{1+2\epsilon^2}\, v \log\left(\frac{\Lambda}{m_H}\right) \;,\;
m_{\mu}^{(1)}\simeq \frac{\alpha_Y^{32}\epsilon_2 }{16\pi^2} \sqrt{1+\epsilon^2}\, v\log\left(\frac{\Lambda}{m_H}\right)\,,
\end{align}
from the mass matrix $\bM_\ell^{(1)} = v \bY_3^{\ell (1)}$. Taking $(\epsilon_1,\epsilon_2,\epsilon_3)\simeq (1.5\times 10^{-5},3.2\times 10^{-3},0.05)$ and $\epsilon \ll 1$, while keeping $\alpha_Y^{3i}\sim \mathcal{O}(1)$, allows one to reproduce the observed values for $m_{e,\mu,\tau}$. Moreover, the left-handed rotation which brings $\bM_\ell^{(1)}$ to the diagonal basis is no longer trivial:
\begin{equation}
\bV_L\simeq\left(
\begin{array}{ccc}
-1+{\epsilon^2} &{\epsilon}   &{\epsilon}  \\
{\epsilon} &1- \dfrac{\epsilon^2}{2}&0 \\
{\epsilon}  &-{\epsilon^2}  &1- \dfrac{\epsilon^2}{2}
\end{array}
\right)\,.
\end{equation}
Such a rotation would be invaluable to correct, for instance, a tribimaximal (TBM) mixing pattern~\cite{Harrison:2002er} arising at tree level, as its exact version is presently excluded due to a nonzero reactor neutrino angle. Were this the case in the above example, one would have the correction $\bU_{\rm TBM} \rightarrow \bV_L^\dag\bU_{\rm TBM}$ to the lepton mixing matrix due to the rotation of charged-lepton left-handed fields to the diagonal basis. The new mixing angles would then be:
\begin{align}
\label{mixang}
\sin^2\theta_{12} \simeq \frac{1}{3}+\frac{4}{9}\epsilon\;,
\;\sin^2\theta_{23} \simeq \frac{1}{2}(1-{\epsilon^2})\;,\;
\sin^2\theta_{13}\simeq  2\epsilon^2\,,
\end{align}
which, for $\epsilon \simeq -0.11$, lead to the right value for the reactor neutrino angle $\theta_{13}$, while keeping the remaining mixing angles within experimentally allowed ranges~\cite{Tortola:2012te}. This simple example shows how a scenario which would be excluded by tree-level considerations becomes phenomenologically viable when quantum corrections to the charged-lepton Yukawa couplings are included. We point out that we have started from a situation where the muon and the electron were massless and $\theta_{13}=0$, to end up with a case where $m_e$, $m_\mu$ and $\theta_{13}$ are radiatively generated. Quark masses and mixing can also be generated through this same effect~\cite{Ibarra:2014fla}.

Although we have presented examples with massless charged leptons at tree level---which would certainly call for a justification---we stress that, even if this is not the case, the RG corrections to the Yukawas in NHDMs should always be kept in mind. To illustrate this, consider a 2HDM with real and diagonal Yukawa matrices $\bY_a^\ell={\rm diag}(y_{a1},y_{a2},y_{a3})$ in the Higgs basis where $\langle \phi_2^0 \rangle=0$. The one-loop corrected Yukawas will be
\begin{equation}
\label{yis}
y_{1i}^{(1)} \simeq y_{1i} - \frac{(\delta_{1b}\alpha_g+\alpha_Y^{1b})}{16 \pi^2}y_{bi} \log\left(\dfrac{\Lambda}{m_H}\right)\,.
\end{equation}
Taking the natural value $\alpha_Y^{12} \sim \mathcal{O}(1)$, it is apparent from the above estimate that if $y_{2i} \gtrsim 16\pi^2 y_{1i}/$ $\log(\Lambda/m_H)$, the one-loop contributions to the Yukawa couplings are relevant. If, for instance, the electron couples with $\Phi_1$ and $\Phi_2$ with $y_{1e} = m_e/v \simeq 2.9 \times 10^{-6} $ and $y_{2e} \gtrsim 4.6\times 10^{-4}/\log(\Lambda/m_H)$, the correction to its coupling will be of the same order (or larger) than the tree-level value. Choosing as before $\Lambda\sim 10^{16}$~GeV  and $m_H \sim 1$~TeV, this bound becomes $y_{2e} \gtrsim 3.5\times 10^{-5}$, which may still be a relatively small parameter. For the third generation, $y_{2\tau} \gtrsim 0.1$ would lead to a correction of the order of $y_{1\tau}=m_\tau/v$. In fact, even if each charged lepton couples with the same strength with the two Higgs doublets, the Yukawa couplings would become $y_{1i}^{(1)}=y_{1i} -\delta y_{1i}$, where
\begin{equation}
\label{yisratio}
\frac{\delta y_{1i}}{y_{1i}}\simeq  \frac{(\alpha_g+\alpha_Y^{11}+\alpha_{Y}^{12})}{16 \pi^2}\log\left(\dfrac{\Lambda}{m_H}\right)\,.
\end{equation}
Taking $\alpha_g+\alpha_Y^{11}+\alpha_{Y}^{12} \simeq  4$, $\Lambda\sim 10^{16}$~GeV, and $m_H \sim 1$~TeV, the induced $\delta y_{1i}$ amounts to $\sim 70\%$ of the tree-level masses.

It is worth pointing out that loop effects not controlled by the RGE running, though out of the scope of the present analysis, may need to be considered. These contributions have recently been shown to be at most 5\% in several versions of the 2HDM \cite{Kanemura:2014dja}.

\section{Radiative charged-lepton masses in NHDM\lowercase{s} with right-handed neutrinos}
Consider now the possibility of extending an NHDM by adding three right-handed neutrinos $\nu_{Rj}$ to the field content, with masses $M_j \gg v$. The extended Yukawa Lagrangian reads 
\begin{equation}
\label{Lqlnu}
\mathcal{L}^\prime=\mathcal{L}-\left[(\bY_a^\nu)_{ij} \bar{\ell}_{Li} \tilde{\Phi}_a \nu_{Rj} +{\rm H.c.}\right]\,,
\end{equation}
where $\mathcal{L}$ is that of Eq.~(\ref{Lql}), while Dirac neutrino Yukawa matrices are denoted by $\bY_a^\nu$. 

It is known that in the SM the presence of couplings $\bY^\nu$ cannot generate charged-lepton masses radiatively. This is due to the fact that new terms in the beta function of $\bY^\ell$ are of the form $\bY^\nu \bY^{\nu\dag} \bY^\ell$, implying that Yukawa eigenvalues are always proportional to themselves~\cite{Antusch:2005gp}. In contrast, different $\bY_k^\nu$ in the NHDM will contribute to the renormalization of the various $\bY_j^\ell$, both through wave-function and vertex corrections above the right-handed neutrino mass scale. It can be shown that the modified one-loop beta function, valid from $\Lambda > M_i$ to $M_i$, reads
\begin{equation}
\label{betanu}
\beta_{a}^{\prime(1)}=\beta_{a}^{(1)}+\alpha_\nu^{ab} \bY_b^\ell+\frac{1}{2} \bY_b^\nu \bY_b^{\nu\dag} \bY_a^\ell   -2\bY_b^\nu\bY_a^{\nu\dag}\bY_b^\ell\,,
\end{equation}
with $\alpha_\nu^{ab}={\rm Tr}(\bY_a^{\nu\dag} \bY_b^\nu)$, in agreement with Ref.~\cite{Ibarra:2011gn}. The presence of right-handed neutrinos adds two new contributions to the beta function $\beta_a^{(1)}$ which do not depend on $\bY_a^\ell$ [cf.~Eqs.~\eqref{beta}~and~\eqref{betanu}]. Notice, however, that these extra terms are only active for $\mu \in [M_i,\Lambda>M_i]$. 

It can be seen that the second term of Eq.~(\ref{betanu}) alone can generate masses for both the electron and the muon in an NHDM. Extending the previously presented 3HDM example by the addition of right-handed neutrinos, and considering, for simplicity, diagonal Yukawa matrices in the Higgs basis, one obtains
\begin{equation}
\label{massneut}
m_{e_{i}}^{(1)}\simeq \frac{v}{16 \pi^2}\alpha_\nu^{21} \epsilon_{i} \log\left( \frac{\Lambda}{M_i}\right)\;,\; i=1,2\,,
\end{equation}
where $\epsilon_{i}$ are the couplings of the electron ($i=1$) and the muon ($i=2$) to the zero-VEV Higgs doublet $\Phi_1$.
Taking $\bY_k^\nu\sim \mathcal{O}(1)$, which implies $\alpha_\nu^{21} \sim \mathcal{O}(1)$, seesaw (light) neutrino masses $m_\nu \sim 0.05$~eV require a heavy scale of $M_i\sim 10^{14}$~GeV. In this case, one could obtain the right electron and muon masses for $(\epsilon_1,\epsilon_2) \simeq (10^{-4}, 0.02)$. If, on the other hand, one assumes that charged leptons couple with the same strength to the doublets $\Phi_{1}$ and $\Phi_{2}$, the corrections to the masses coming from the $\bY_k^\nu$'s do not exceed $10\%$ of the tree-level values.

\section{Conclusions}
In this work it has been shown that charged-lepton masses can be radiatively generated in a natural way in NHDMs due to the presence of Yukawa couplings of leptons and quarks with the extra (non-SM) Higgs doublets. It is important to mention that, besides affecting charged-lepton masses, the RGE running may have dramatic effects on the left-handed rotation which brings the charged-lepton mass matrix to its diagonal form. This clearly changes predictions for lepton mixing parameters, which have to be confronted with neutrino oscillation data. These corrections should then be taken into account in phenomenological studies of fermion mass and mixing models with more than one Higgs doublet. The presence of right-handed neutrinos provides additional sources for this effect.
\acknowledgments
This work has been supported by Projects No. EXPL/FIS-NUC/0460/2013, No. CERN/FP/123580/2011 and No. \mbox{PEst-OE/FIS/UI0777/2013}, financed by \textit{Funda\c c\~ao para a Ci\^encia e a Tecnologia} (FCT, Portugal).

\end{document}